\documentstyle[aps,preprint]{revtex}

\begin{document}

\draft
\title
{
Elementary Excitations and Dynamical Correlation Functions \\
of the Calogero-Sutherland Model with Internal Symmetry
}
\author
{ 
Yusuke {\sc Kato}\footnote{kato@coral.t.u-tokyo.ac.jp}, 
Takashi {\sc Yamamoto}$^{1,}$\footnote{yam@yukawa.kyoto-u.ac.jp}
and Mitsuhiro {\sc Arikawa}$^{2}$
}
\address{
Department of Applied Physics, University of Tokyo, Tokyo 113\\
$^1$Yukawa Institute for Theoretical Physics, Kyoto University, Kyoto 606\\
$^2$Department of Physics, Tohoku University, Sendai 980-77
}
\date{Mar. 7, 1997}
\maketitle
\begin{abstract}
We consider the physical properties of elementary excitations of 
the Calogero-Sutherland (CS) model with SU($K$) internal symmetry. 
{}From the results on the thermodynamics of this model, 
we obtain the charge, spin, and statistics of elementary excitations. 
Combining this knowledge and the known results 
on the dynamics in the spinless CS model, 
we propose the expression for the dynamical correlation functions 
of the SU($K$) CS model. 
In the asymptotic region, we confirm the consistency of our results 
with predictions from conformal field theory. 
\end{abstract}
%
%
%
%

\newpage

\section{Introduction}

Over the past years, progress has been made with the study 
of the Calogero-Sutherland (CS) model\cite{Cal,Suth} which describes 
a one-dimensional system of particles 
interacting with inverse-square interactions.
It has turned out that this model has a remarkably simple structure,
and can be considered as a model of ideal gas 
obeying fractional statistics.\cite{Haldanefrac}
Furthermore, exact results are now available for some ground state 
dynamical correlation functions.\cite{%
SLA,HZ,Ha,LPS,Forr,Konn,MP1,Khve,Tsve}
Here we shall briefly recall these works.

A breakthrough was brought by Simons et al.\cite{SLA}
They found that the dynamical density-density correlation
functions of the CS model at couplings 
$\beta=1/2$ and $2$ (see the Hamiltonian (\ref{hamil}))
are identical to some parametric correlators in the random systems.
The latter can be obtained in terms of the so-called supermatrix method.
Using the similar method, Haldane and Zirnbauer calculated
the one-particle retarded dynamical Green function at $\beta=2$.\cite{HZ} 
{}From these results together with the knowledge of noninteracting 
case ($\beta=1$), Haldane then conjectured the expression of 
the dynamical correlation functions 
at arbitrary rational couplings.\cite{Hal-rev}
His argument is based on the consideration of the selection rule
for the intermediate states. Subsequently, this conjecture was 
proved by using the Jack polynomial techniques.\cite{Ha,LPS,Forr}

Among a number of generalizations of the original CS model,
the CS model with SU($K$) internal symmetry
(SU($K$) CS model)\cite{HaHaldane,MP2,HW}
has been attaching much attention in recent years.
It is well known that the lattice variants of the CS model such as 
the Haldane-Shastry model\cite{HS1,HS2} 
and supersymmetric $1/r^2$ $t$-$J$ model\cite{KY} 
can be obtained from the CS model 
with appropriate internal symmetry
in the strong coupling limit.\cite{Polychronakos,SS,KaKu1}
It is natural to examine the dynamical properties of the SU($K$) CS model.
Quite recently, one of the authors obtained (hole part of) the 
dynamical Green function of the SU(2) CS model for 
a special coupling.\cite{Green}
As is the spinless case, the resultant formula of the correlation function
is so simple that he can conjecture 
the expression for arbitrary integer couplings.

In this paper, we systematically study the elementary excitations 
of the SU($K$) CS model from the result obtained by the thermodynamics.
Then, we propose the expression for the dynamical
correlation functions of the SU($K$) CS model. 
Our approach is close to ref. \cite{Hal-rev} in spirit.

After finishing our calculations, 
we come to know that Uglov has performed rigorous calculations 
for the dynamical density and spin-density correlation functions 
of the SU(2) CS model (without taking the thermodynamic limit).\cite{Uglov}
Since it is sometimes useful to study complicated problems from
several different point of view,
we believe it still makes sense to present the details of our 
approach and of our arguments.

The content of this paper is as follows.
In section 2, we recall the thermodynamics and dynamics 
of the spinless CS model. Then, following ref. \cite{Hal-rev},
we give the interpretation of the formula for 
the dynamical density-density correlation function
in terms of the knowledge about the elementary excitations.
In section 3, we extend the argument in section 2
to the SU($2$) case. We also perform the expansion of correlation function
and then check the consistency with the predictions from 
conformal field theory.
In section 4, we generalize to the  SU($K$) case ($K>2$).
Finally, in section 5, we summarize and discuss our results.

\section{Review on spinless Calogero-Sutherland model}

Before considering the model with internal symmetry, 
we review the explicit results on thermodynamics\cite{Suth} 
and dynamics\cite{SLA,HZ,Ha,LPS,Forr,Konn,MP1,Khve,Tsve,Hal-rev}
of the spinless model. 
In this section, first we derive the one-particle energy, momentum,
charge and exclusion statistics of elementary excitations from 
the thermodynamics. Next we show that known results 
on the dynamics can be interpreted in a simple manner 
in terms of the elementary excitations derived from the thermodynamics
(see also ref. \cite{Hal-rev}).

We consider the following Hamiltonian: 
\begin{equation}
\hat{\cal H}
=
-\frac12\sum_{i=1}^N \frac{\partial ^2}{\partial x_i^2}
+\frac{\pi^2}{L^2}\sum_{i<j}
\frac{\beta\left(\beta-1\right)}
     {\sin^2\left[\pi\left(x_i-x_j\right)/L \right]}
\label{hamil}
\end{equation}
for $N$-particle system on a circle with the linear dimension $L$. 
We restrict attention to the case that 
the dimensionless coupling parameter 
$\beta$ takes non-negative integer values,
although more general results are now obtained.\cite{Ha} 
The statistics of particles are chosen 
as boson (fermion) for even (odd) $\beta$. 

Adopting the Yang and Yang's method,\cite{YY}
Sutherland constructed the thermodynamics of spinless CS model.\cite{Suth}
The thermodynamic potential (per unit length) is given by 
\begin{equation}
\Omega
=
-T\int_{-\infty}^{\infty}\frac{{\rm d}v}{2\pi}
\ln\left(1+\omega^{-1}\right),
\end{equation} 
where $\omega=\omega(v,T)$ is the real solution 
of the following equation:
\begin{equation}
\label{realsolution}
\left(v^2/2-\zeta\right)/T
=\log\left(1+\omega\right)
-\beta\log\left(1+\omega^{-1}\right), 
\end{equation}
with the chemical potential $\zeta$. 
We call the integration variable $v$ velocity, 
in terms of which the CS model is described as a free particle system. 
For the spinless model, thermodynamics was reformulated 
in terms of free particles obeying 
the fractional exclusion statistics.\cite{Haldanefrac}
In the statistical mechanics of 
the fractional exclusion statistics\cite{Wu,BW}, 
the velocity distribution function of particles $\rho(v)$, satisfying 
\begin{equation}
\label{rho0}
\int_{-\infty}^{\infty}\frac{{\rm d}v}{2\pi}\rho(v)
=\frac{N}{L}
\equiv \rho_0,
\end{equation}
is given by $\rho(v)=1/(\omega+\beta)$. 
Also, the distribution function of hole or vacant one-particle state 
$\rho^*(v)$ is given by $\rho^* (v)=\omega/(\omega+\beta)$. 
Thus, $\omega$ represents the ratio of the distribution of holes 
to that of particles. 

Now let us rewrite the thermodynamics in terms of 
{\it elementary excitations}. 
For this purpose, we first consider the ground state. 
At $T=0$, the distributions of particles and holes are given by
\begin{equation}
\label{rhop}
\left\{
\begin{array}{lll}
\rho(v)=1/\beta,       &
\rho^*(v)=0,   \quad &
\mbox{for}\ \ 
\left|v\right|<v_{\rm F}\equiv\left(2\zeta\right)^{1/2},\\
\rho(v)=0,           &
\rho^*(v)=1,   \quad &
\mbox{for}\ \ 
\left|v\right|>v_{\rm F}.
\end{array}
\right.
\end{equation}
Combining eqs. (\ref{rho0}) and (\ref{rhop}), 
we obtain $\zeta=v_{\rm F}^2/2=\left(\pi\rho_0\beta\right)^2/2$. 
{}From the above solution (\ref{rhop}), 
we find that the ground state is analogous to the Fermi sea
and that there are two branches of excitations; 
particle-type for $\left|v\right|>v_{\rm F}$ 
and hole-type for $\left|v\right|<v_{\rm F}$. 
{}From now on, we call the former quasiparticles and the latter quasiholes. 
With the above identification, 
we introduce $\omega_{\rm p}=\omega$ for $\left|v\right|>v_{\rm F}$ 
and $\omega_{\rm h} =\omega^{-1}$ for $\left|v\right|<v_{\rm F}$. 
In terms of $\omega_{\rm p}$ and $\omega_{\rm h}$, 
the equation (\ref{realsolution}) is rewritten as
\begin{equation}
\epsilon_{\rm p}(v)/T
\equiv\left(v^2/2-\zeta\right)/T
=\log\left(1+\omega_{\rm p}\right)
-g_{\rm p}\log\left(1+\omega_{\rm p}^{-1}\right),
\label{omega>}
\end{equation}
and
\begin{equation}
\epsilon_{\rm h}(v)/T
\equiv\left(\zeta-v^2/2\right)/(\beta T)
=\log\left(1+\omega_{\rm h}\right)
-g_{\rm h}\log\left(1+\omega_{\rm h}^{-1}\right),  
\label{omega<}
\end{equation}
with $g_{\rm p}=\beta$ and $g_{\rm h}=1/\beta$. 
Equations (\ref{omega>}) and (\ref{omega<}) respectively have the form of the
equations of the thermodynamics for free particles 
with energy $\epsilon_{\rm p}$, $\epsilon_{\rm h}$ 
and exclusion statistics $g_{\rm p}$, $g_{\rm h}$. 

With the use of eqs. (\ref{omega>}), (\ref{omega<}), 
the thermodynamic potential is rewritten as
\begin{equation}
\Omega
=-T\int_{\left|v\right|>v_{\rm F}}\frac{{\rm d}v}{2\pi}
\ln\left(1+\omega_{\rm p}^{-1}\right)
-\frac{T}{\beta} \int_{\left|v\right|<v_{\rm F}}
\frac{{\rm d}v}{2\pi}\ln\left(1+\omega_{\rm h}^{-1}\right)
-\frac{v_{\rm F}^3}{3\pi \beta}.
\end{equation}
In the above equation, the last term of the right-hand side represents 
the ground state contribution. The first and second terms represent 
the contributions of quasiparticles and quasiholes, respectively. 

The characters of elementary excitations 
are indexed by the charge, statistics, and velocity. 
The (exclusion) statistics are already obtained. 
The charge can be determined by the coefficient of the chemical potential 
in the one-particle energy such as $\epsilon_{\rm p}$ and $\epsilon_{\rm h}$. 
The charge ${\rm e}_{\rm p}$ of quasiparticles is unity 
since the coefficient of $\zeta$ in $\epsilon_{\rm p}$ is $-1$. 
Quasiholes, on the other hand, 
have the fractional charge ${\rm e}_{\rm h}=-1/\beta$ 
since the coefficient of $\zeta$ in $\epsilon _{\rm h}$ is $1/\beta$. 
Momenta of quasiparticles $q_{\rm p}$ and quasiholes $q_{\rm h}$ are given by 
\begin{equation}
q_{\rm p}
=\frac{\partial \epsilon_{\rm p}(v)}{\partial v}=v,\quad 
q_{\rm h}
=\frac{\partial \epsilon_{\rm h}(v)}{\partial v}=-v/\beta. 
\end{equation}
If we set $\beta=1$, which is the free fermion case, 
all the above relations reduce to the trivial one. 

With the knowledge about the elementary excitations, 
we try to interpret the known results of 
the dynamical density-density correlation 
functions.\cite{SLA,HZ,Ha,LPS,Forr,Konn,MP1,Khve,Tsve,Hal-rev} 
The local density operator excites only neutral (charge zero) excitations. 
Minimal process of neutral excitations are one quasiparticle 
and $\beta$ quasiholes excitation, which we call ^^ ^^ minimal bubble". 
(Recall that the coupling constant $\beta$ takes values 
in non-negative integers.)
The charge neutrality is confirmed by the relation 
${\rm e}_{\rm p}+\beta {\rm e}_{\rm h}=0$. 
The intermediate states can be expanded by the states 
corresponds to the multiple excitation of minimal bubble. 

Now we show that the dynamics can be interpreted 
with the above knowledge of the elementary excitations.
The expression of the dynamical density-density correlation function 
is given by\cite{SLA,HZ,Ha,LPS,Forr,Konn,MP1,Khve,Tsve,Hal-rev} as
\begin{equation}
\langle\rho(x,t)\rho\rangle 
\propto 
\int_{|u| \geq v_{\rm F}} {\rm d} u
\prod_{j=1}^{\beta}\int_{|v_j| \leq v_{\rm F}} {\rm d} v_j 
\exp[ {\rm i}(Q x - E t) ] 
Q^2F(u,\left\{v_i\right\}), 
\label{rhorholess}
\end{equation}
where
\begin{equation}
\label{eq}
Q=u-\frac1\beta\sum_{j=1}^{\beta} v_j, \quad
E=\epsilon_{\rm p}(u)+\sum_{j=1}^{\beta} \epsilon_{\rm h}(v_j), 
\end{equation}
and 
\begin{equation}
F(u,\left\{v_i\right\})
=\frac{\prod_{j=1}^{\beta} ( u - v_j )^{-2}
       \prod_{i < j}^{\beta} (v_i - v_j)^{2 g_{\rm h}}     }
      {\left[ \epsilon_{\rm p} (u)\right]^{1- g_{\rm p} }
       \prod_{j=1}^{\beta}
       \left[ \epsilon_{\rm h} (v_j)\right]^{1-g_{\rm h}}  }. 
\end{equation}

First we set aside the interpretation of the form factor 
$ Q^2 F(u,\left\{v_i\right\})$ 
and consider the rest of expression (\ref{rhorholess}). 
{}From the integral regions in (\ref{rhorholess}), 
we are led to the interpretation that $u$ is the velocity of quasiparticle 
and $v_i$ are those of quasiholes. 
It means that the local density operator $\rho(x)$ excites 
{\it only} the minimal bubble in the CS model.
According to expression (\ref{eq}), 
excited quasiparticle and quasiholes are energetically free. 
Expression (\ref{eq}) is consistent with the dispersion relation 
obtained from the thermodynamics.

Next we consider $ F(u,\left\{v_i\right\})$, 
which gives a nontrivial part of the form factor. 
In the thermodynamics, the system is energetically free 
but has the nontrivial structure brought by 
the statistical interactions\cite{Haldanefrac} among elementary excitations. 
Hence we try interpreting each factor of $F(u,\left\{v_i\right\})$
in terms of statistical interactions as follows:
\begin{itemize}
\item 
The factor $\left(v_i-v_j\right)^{2g_{\rm h}}$ per each pair 
is caused by the statistical interactions among quasiholes. 
Since the exponent $2g_{\rm h}$ is fractional, 
we attribute the statistical interaction to the two-body interaction 
in (\ref{hamil}). 

\item 
The factor $\left(u - v_j\right)^{-2}$ comes from the statistical interactions 
between quasiparticle and quasiholes.
Since the exponent is independent of $\beta$, 
we attribute the statistical interaction to the original Pauli exclusion. 

\item 
The factor $ [\epsilon(u)]^{g_{\rm p}-1}$ 
comes from the statistical interaction of excited quasiparticle with itself.

\item 
The factor $ [\epsilon(v_i)]^{g_{\rm h}-1}$ comes from each excited quasihole. 
These additional factors stem from the statistical interaction 
of each excitation with itself. 
\end{itemize}

Finally we consider the factor $Q^2$. 
{}From the process of the microscopic calculation, 
we know that the factor $Q^2$ is due to the local density operator 
$\rho(x)=\sum_{i=1}^N \delta(x-x_i)-\rho_0$. 

{}From the above observation, we learn that the dynamics can be interpreted 
in terms of the knowledge of elementary excitations 
derived from the thermodynamics. 
Especially, we emphasize that
the nontrivial form factor is determined by 
the statistical interactions and one-particle energy of excitations. 

For the SU($K$) CS model, 
thermodynamics has been formulated\cite{fesmsm} in a similar way 
with the spinless case. 
Hence we can obtain the physical properties of elementary excitations 
such as dispersion, charge, statistics, and spin (or color).
In the next section we will assume that the above observation holds 
even in the SU(2) CS model 
and construct the dynamical correlation function
in a similar fashion as the spinless case.

\section{Calogero-Sutherland model with SU(2) internal symmetry}

We consider the system of particles with SU(2) spin 
$\sigma\in\{+1, -1\}$, 
whose Hamiltonian is given by

\begin{equation}
\hat{\cal H}
=
-\frac12\sum_{i=1}^N \frac{\partial ^2}{\partial x_i^2}
+\frac{\pi^2}{L^2}\sum_{i<j}
\frac{\beta\left(\beta+P_{ij}\right)}
     {\sin^2\left[\pi\left(x_i-x_j\right)/L \right]}.
\label{hamilmulti}
\end{equation}
Here $\beta$ takes non-negative integer values, and 
$P_{ij}$ denotes the operator that exchanges 
the spin of particles $i$ and $j$. 
The statistics of particles are chosen 
as boson (fermion) for odd (even) $\beta$.  
Notice that the spinless Hamiltonian (\ref{hamil}) 
can be recovered from (\ref{hamilmulti}) by taking $P_{ij}\equiv 1$ and 
$\beta\rightarrow\beta-1$.

\subsection{Elementary excitations and dynamical correlation functions}

For this model, the thermodynamic potential 
is give by\cite{fesmsm}
\begin{equation}
\Omega
=-T\int_{-\infty}^\infty \frac{{\rm d}v}{2\pi}
\sum_{\sigma=\pm 1}\ln\left(1+\omega_{\sigma} ^{-1}\right),
\label{Omega2}
\end{equation}
where $\omega_{\sigma}$ is the real solution of the following equations:
\begin{equation}
\epsilon_{{\rm p}\sigma}(v)/T
\equiv\left(v^2/2-\zeta-\sigma h\right)/T
=\ln\left(1+\omega_\sigma\right)
-\sum_{\sigma'=\pm 1}g_{\rm p}^{\sigma,\sigma'}
\ln\left(1+\omega_{\sigma'}^{-1}\right), \ \ 
(\sigma=\pm 1).
\label{epsu2}
\end{equation}
Here $h$ represents the external magnetic field. 
The matrix ${\bf g}_{\rm p}$ is given by\cite{fesmsm,FK1}
\begin{equation}
\label{gp}
{\bf g}_{\rm p} 
=
\left(
\begin{array}{cc}
g_{\rm p}^{\uparrow \uparrow}  & g_{\rm p}^{\uparrow \downarrow} \\
g_{\rm p}^{\downarrow \uparrow}& g_{\rm p}^{\downarrow \downarrow} 
\end{array}
\right)
=
\left(
\begin{array}{cc}
\beta+1 & \beta  \\
\beta   & \beta+1 
\end{array}
\right),
\end{equation}
which represents the statistics of particles. 
Here we used the notation $\uparrow, \downarrow$ instead of $+1, -1$. 
Equations (\ref{epsu2}) and (\ref{gp}) determine the thermal
distribution of free particles 
with energy $\epsilon_{{\rm p}\sigma}$ and 
exclusion statistics ${\bf g}_{\rm p}$. 
There is a duality representation of eq. (\ref{epsu2}); multiplying it by  
\begin{equation}
{\bf g}_{\rm h}
\equiv
\left(
\begin{array}{cc}
g_{\rm h}^{\uparrow \uparrow}  & g_{\rm h}^{\uparrow \downarrow} \\
g_{\rm h}^{\downarrow \uparrow}& g_{\rm h}^{\downarrow \downarrow} 
\end{array}
\right)
=
{\bf g}_{\rm p}^{-1}
=
\frac{1}{2\beta+1}
\left(
\begin{array}{cc}
\beta+1 & -\beta\\
-\beta  & \beta+1 
\end{array}
\right), 
\end{equation}
we obtain 
\begin{equation}
\epsilon_{{\rm h}\sigma}(v)/T
=\ln\left(1+\omega_{{\rm h}\sigma }\right)
-\sum_{\sigma'=\pm 1}g_{\rm h}^{\sigma \sigma'}
\ln\left(1+\omega_{{\rm h}\sigma'}^{-1}\right), 
\end{equation}
with $\epsilon_{{\rm h}\sigma}(v)
=(\zeta -v^2/2)/(2\beta+1)+\sigma h$ 
and $\omega_{{\rm h}\sigma}=\omega^{-1}_{\sigma}$.

Now we consider excitations from the ground state in the unpolarized
case ($h=0$). 
At $T=0$, the distributions of particles and holes are given as
\begin{equation}
\left\{
\begin{array}{lll}
\rho_\sigma(v)=1/(2\beta +1), &
\rho^*_{\sigma}(v)=0,   \quad &
\mbox{for}\ \ 
\left|v\right|<v_{\rm F}\equiv\left(2\zeta\right)^{1/2},\\
\rho_{\sigma}(v)=0,           &
\rho^*_{\sigma}(v)=1,   \quad &
\mbox{for}\ \ 
\left|v\right|>v_{\rm F}.
\end{array}
\right.
\end{equation}
The chemical potential $\zeta$ is obtained with the relation
\begin{equation}
\int_{-\infty}^{\infty}\frac{{\rm d}v}{2\pi}
\sum_{\sigma=\pm 1}\rho_{\sigma}\left(v\right)
=\frac{N}{L}\equiv \rho_0
\end{equation}
as $\zeta=v_{\rm F}^2/2=\left[\pi \rho_0 \left(2\beta+1\right)/2\right]^2/2$. 

For $\left|v\right|>v_{\rm F}$, 
excitations are particle-like; quasiparticles with energy 
$\epsilon_{{\rm p}\sigma }$, charge ${\rm e}_{\rm p}=+1$, 
spin $\sigma_{\rm p}=-\sigma $, and statistics ${\bf g}_{\rm p}$. 
For $\left|v\right|<v_{\rm F}$, on the other hand, 
excitations are hole-like; quasiholes with energy 
$\epsilon_{{\rm h}\sigma}$, charge $e_{\rm h}=-1/(2\beta+1)$, 
spin $\sigma_{\rm h}=\sigma$, and statistics ${\bf g}_{{\rm h}}$. 
Here we note that charge of quasiholes is renormalized to be fractional 
while spin of those is the same as that of quasiparticles. 
In table \ref{fulltable:1}, 
we summarize above datum about the elementary excitations.

With the use of the knowledge from the thermodynamics, 
we postulate an empirical rule for construction of 
the dynamical density-density correlation function 
$\langle \rho(x,t)\rho\rangle$.
For this correlation function, only the intermediate states 
with charge-zero and spin-zero can contribute. 
These excitations are expressed by 
multiple set of the following minimal bubble:
\begin{equation}
\left\{
\begin{array}{l}
\mbox{one quasiparticle with }   \sigma_{\rm p}=\sigma\\
\beta+1 \mbox{ quasiholes with } \sigma_{\rm h}=-\sigma\\
\beta \mbox{ quasiholes with }   \sigma_{\rm h}=\sigma.
\end{array}
\right.
\end{equation}
The charge and spin neutralities are verified by
\begin{equation}
{\rm e}_{\rm p}+\left(\beta+1\right){\rm e}_{\rm h}+\beta e_{\rm h}=0
\end{equation}
and
\begin{equation}
\sigma+\left(\beta+1\right)\left(-\sigma\right)+\beta \sigma=0,
\end{equation}
respectively. 
{}From the observation of spinless case, we {\it assume} 
that only the minimal bubble contributes to 
the dynamical density-density correlation function.

With this assumption, we write down 
the dynamical density-density correlation function in the following form:
\begin{equation}
\label{dddcf-2}
\langle\rho(x,t)\rho\rangle
=
C_{\rm c}
\int_{\left|u\right|>v_{\rm F}}{\rm d}u\prod_{i=1}^{\beta +1}
\int_{\left|v_i\right|<v_{\rm F}}{\rm d}v_i\prod_{j=1}^{\beta}
\int_{\left|w_j\right|<v_{\rm F}}{\rm d}w_j
\exp[{\rm i}\left(Qx -Et\right)]
Q^2F(u,\left\{v_i\right\},\left\{w_j\right\}),
\end{equation} 
with an unknown constant $C_{\rm c}$, which is considered later. 
Here $E$ and $Q$ are given by
\begin{eqnarray}
E&=&\epsilon_{\rm p}(u)
+\sum_{i=1}^{\beta+1}\epsilon_{\rm h}(v_i)
+\sum_{j=1}^\beta \epsilon_{\rm h}(w_j),
\\
Q&=&u
-\frac{1}{2\beta+1}\left(\sum_{i=1}^{\beta+1}v_i+\sum_{j=1}^\beta w_j\right). 
\end{eqnarray}
The variables $u$, $v_i$ and $w_j$ represent 
the velocities of quasiparticle with spin $\sigma$, and quasiholes 
with spin $-\sigma$ and $\sigma$, respectively. 
Now we consider the integrand
$F(u,\left\{v_i\right\},\left\{w_j\right\})$. 
With the knowledge obtained in the previous section, 
we construct  $F(u,\left\{v_i\right\},\left\{w_j\right\})$ from 
the following factors:
\begin{itemize}
\item Statistical interactions among quasiholes\\
$ \prod_{i < j}^{\beta +1} (v_i - v_j)^{2 g_{\rm h}^{\uparrow \uparrow}}$, 
$ \prod_{i < j}^{\beta} (w_i - w_j)^{2 g_{\rm h}^{\downarrow \downarrow}}$,   
$ \prod_{i=1}^{\beta+1}\prod_{j=1}^{\beta} 
          (v_i - w_j)^{2 g_{\rm h}^{\uparrow \downarrow}}$. \\
Here the exponents 
$g_{\rm h}^{\uparrow\uparrow}=g_{\rm h}^{\downarrow \downarrow}
=\left(\beta+1\right)/(2\beta+1)$ 
and $g_{\rm h}^{\uparrow \downarrow}=-\beta/(2\beta+1)$ 
represent the statistical interactions between quasiholes with 
same spin and opposite spin, respectively. 
\item Statistical interactions between quasiparticle and quasiholes\\
      $ \prod_{j=1}^{\beta+1} 
         (u - v_j)^{-2 \delta_{\sigma_{\rm p}, -\sigma_{\rm h}}} $.\\
Here, $\delta_{\sigma,\sigma'}$ represents the Kronecker's delta. 
\item per quasiparticle\\
      $ ( \epsilon_{\rm p} )^{-1+g_{\rm p}^{\sigma,\sigma}} $.
\item per quasiholes\\
      $ ( \epsilon_{\rm h} )^{-1+g_{\rm h}^{\sigma,\sigma}} $.
\end{itemize}
By putting  all the above together, 
we obtain the following expression for the integrand $F$:
\begin{eqnarray}
F(u,\left\{v_i\right\},\left\{w_j\right\})
=
\frac{
\prod_{i < j}^{\beta +1}
 (v_i - v_j )^{2 g_{\rm h}^{\uparrow \uparrow}}
\prod_{i < j}^{\beta}
 (w_i - w_j )^{2 g_{\rm h}^{\downarrow \downarrow}}
\prod_{i=1}^{\beta+1}\prod_{j=1}^{\beta}
 (v_i - w_j)^{2 g_{\rm h}^{\uparrow \downarrow}}                  }
     {
\prod_{i=1}^{\beta+1}(u - v_i)^2
[\; \epsilon_{\rm p}(u)  \; ]^{1-g_{\rm p}^{\uparrow \uparrow}}
[\; \prod_{i=1}^{\beta+1}
    \epsilon_{\rm h}(v_i)\; ]^{1-g_{\rm h}^{\uparrow \uparrow}}
[\; \prod_{j=1}^{\beta}
    \epsilon_{\rm h}(w_j)\; ]^{1-g_{\rm h}^{\downarrow \downarrow}}}.
\label{rhorhoarrive}
\end{eqnarray}

The unknown factor $C_{\rm c}$ in eq. (\ref{dddcf-2}) 
is obtained by the known static charge susceptibility:
\begin{equation}
\chi_{\rm c}
=\left(\frac{\partial^2 E}{\partial N^2}\right)^{-1}
=\frac{4}{\pi^2 \rho_{0}\left(2\beta+1\right)^2} 
\end{equation}
and the Kramers-Kronig relation:
\begin{equation}
\chi_{\rm c}
=
\lim_{q \rightarrow 0}\frac{2}{\pi}
\int_{0}^{\infty}\frac{{\rm d}\omega}{\omega}
\left[
\int_{-\infty}^{\infty}{\rm d}x
\int_{-\infty}^{\infty}{\rm d}t 
\langle\rho(x,t)\rho\rangle\exp\left[{\rm i}\left(\omega t-qx\right)\right]
\right]. 
\end{equation}
{}From the above equation, we obtain 
$C_{\rm c}=\left(4\pi^2\left(2\beta+1\right)^{\beta +1}S(\beta)\right)^{-1}$ 
with 
\begin{eqnarray}
S(\beta)
&=&
\prod_{i=1}^{\beta+1}\int_0 ^1 {\rm d}p_i 
\prod_j^\beta \int_0^1 {\rm d}q_j 
\prod_{i=1}^{\beta+1}
\left[p_i\left(1-p_i\right)\right]^{g_{\rm h}^{\uparrow \downarrow}}
\prod_{j=1}^\beta 
\left[q_j\left(1-q_j\right)\right]^{g_{\rm h}^{\uparrow \downarrow}}
\nonumber\\
& &
\prod_{i<j}^{\beta+1}
\left(p_i -p_j\right)^{2g_{\rm h}^{\uparrow \uparrow}}
\prod_{i<j}^{\beta}\left(q_i -q_j\right)^{2g_{\rm h}^{\uparrow \uparrow}}
\prod_{i=1}^{\beta+1}\prod_{j=1}^{\beta}
\left(p_i -q_j\right)^{2g_{\rm h}^{\uparrow \downarrow}}.
\end{eqnarray}

In this way, we arrive at the complete expression of 
$\langle \rho(x,t)\rho\rangle$. 

\subsection{Asymptotic behavior of correlation function}

Next we consider the expression of the above dynamical correlation
function in the asymptotic region 
and examine the consistency with the predictions from 
conformal field theory (CFT). 

As we presented so far, elementary excitations 
are either particle-type or hole-type. In low energy limit, 
on the other hand, excitations are also described 
in terms of collective excitations. 
In this region, CFT predicts 
asymptotic behaviors of correlation functions. 
For the  SU$(K)$ CS model, 
asymptotic behaviors of correlation functions 
have been considered in ref. \cite{Kawakami} with the use of CFT. 
In the rest of this section, we first perform the asymptotic 
expansion of the expression (\ref{rhorhoarrive}) 
and then compare it with the CFT prediction. 

For the long-distance behavior of the expression (\ref{rhorhoarrive}), 
main contribution comes from the excited states where all excitations 
are near the Fermi points; 
$u\sim \pm v_{\rm F}$, $v_i\sim \pm v_{\rm F}$, and $w_j\sim\pm v_{\rm F}$. 
The asymptotic expression is given by the sum of the contributions from 
the ^^ ^^ sectors"; each sector is specified by 
the distribution of excitations in the two Fermi points. 
We denote by the quartet $(\sigma,\pm 1,m,n)$ with 
$m\in\{0, 1, 2, \cdots, \beta+1\}, n\in\{0, 1, 2, \cdots, \beta\}$
the sector where quasiparticle with spin $\sigma$ is near $u=v_{\rm F}$, 
$m$ quasiholes with spin $-\sigma$  near $v_i=\mp v_{\rm F}$ , 
and $n$ quasiholes with spin $\sigma$ near $w_j=\mp v_{\rm F}$. 
By similar calculations in ref. \cite{Ha}, 
we obtain the following expression:

\begin{equation}
\label{rhorhoasym}
\langle \rho \left(x,t\right)  \rho \rangle
= A \left(
      \frac{1}{\xi_{\rm L}^2}+\frac{1}{\xi_{\rm R}^2}
    \right)
+ \sum_{m,n} A_{m n} 
     \left(
       \frac{1}{\xi_{\rm L} \xi_{\rm R}}
     \right)^{g_{\rm h}^{\uparrow \uparrow}(m^2+n^2)
+ 2g_{\rm h}^{\uparrow \downarrow} m n}
     \cos \left[
            2(\pi \rho_0/2)( m +  n) x
          \right], 
\end{equation}
with $ \xi_{\rm L} = x + v_{\rm F}t ,\; 
       \xi_{\rm R} = x - v_{\rm F}t $.
Here $A$ and $A_{mn}$ are 
$x$- and $t$-independent constants. 
The sum is taken over $0\le m\le \beta+1$ and $0\le n\le \beta$ 
except $(m,n)=(0,0)$. 

In low energy region, on the other hand, the SU(2) CS model 
is categorized to be a Tomonaga-Luttinger liquid.\cite{TLL}  
Each Tomonaga-Luttinger liquid with spin is specified with four velocities; 
the current $v^{J}_{\rm c}$ ($v^{J}_{\rm s}$) 
and charge $v^N_{\rm c}$ ($v^N_{\rm s}$) velocities 
for electronic (spin) sector. 
These parameters take the following values 
in the SU(2) CS model\cite{HaHaldane}:
\begin{equation}
v_{\rm c}^J={\rm e}^{2\alpha} v_{\rm F}, 
\quad v_{\rm c}^N={\rm e}^{-2\alpha}v_{\rm F},
\quad v_{\rm s}^J=v_{\rm s}^N=v_{\rm F}, 
\end{equation}
with ^^ ^^ Bogoliubov angle" ${\rm e}^{\alpha}=\left(2\beta+1\right)^{-1/2}$. 
The relation $v_{\rm s}^J=v_{\rm s}^N$ originates from the SU(2) symmetry. 
In the low energy region, the Hamiltonian decouples into the electronic 
and spin sectors, and hence eigenstates are given by the tensor product 
of eigenstates of the two sectors. 
Eigenstates of the energy and momentum in each sector 
are indexed by the set of four numbers; 
$\left\{(N_{\rm c},D_{\rm c},n_{\rm c}^+,n_{\rm c}^{-})\right\}$ 
for electronic sector, 
$\left\{(N_{\rm s},D_{\rm s},n_{\rm s}^+,n_{\rm s}^{-})\right\}$ 
for spin sector, respectively. 
Here $N_{\rm c}$ and $D_{\rm c}$ are integers or half-odd integers 
representing 
the electronic charge-changing and current excitations respectively. 
The non-negative integer $n_{\rm c}^+$ ($n_{\rm c}^-$) represents 
the number of excited boson of the holomorphic (antiholomorphic) part.
For spin sector, $N_{\rm s}$ and $D_{\rm s}$ represent 
the magnetization-changing and spin current excitations respectively. 
The non-negative integer $n_{\rm s}^{+}$ ($n_{\rm s}^-$) 
is the number of excited boson in spin sector.

For the density-density correlation function, 
CFT predicts the following asymptotic behavior:\cite{BPZ,Cardy}
\begin{equation}
\label{rhorhocft}
\langle \rho(x,t)\rho \rangle
\sim
\sum_{\lambda_{\rm c}, \lambda_{\rm s}}
\frac{
A_{\lambda_{\rm c},\lambda_{\rm s}}
\exp\left[{\rm i}2k_{\rm F}D_{\rm c}x\right]
}
{
(x-v_{\rm c} t)^{2\Delta_{\rm c}^{+}(\lambda_{\rm c})}
(x+v_{\rm c} t)^{2\Delta_{\rm c}^{-}(\lambda_{\rm c})}
(x-v_{\rm s} t)^{2\Delta_{\rm s}^{+}(\lambda_{\rm s})}
(x+v_{\rm s} t)^{2\Delta_{\rm s}^{-}(\lambda_{\rm s})}
}, 
\end{equation}
where the sum is taken over 
$\lambda_{\rm c}\in\left\{(N_{\rm c}=0,D_{\rm c},n_{\rm c}^{\pm})\right\}$ 
and 
$\lambda_{\rm s}\in\left\{(N_{\rm s}=0,D_{\rm s},n_{\rm s}^{\pm})\right\}$. 
Here $k_{\rm F}$ is the Fermi wave number, and the quantity 
$\Delta_{\rm c}^{+}(\lambda_{\rm c})$ ($\Delta_{\rm c}^{-}(\lambda_{\rm c})$) 
represents the conformal weight of holomorphic (antiholomorphic) part 
in the electronic sector. The expression of conformal weight is given by
\begin{equation}
\Delta_{\rm c}^{\pm}(\lambda_{\rm c})
=\left(N_{\rm c}{\rm e}^{-\alpha}\pm D_{\rm c}{\rm e}^\alpha\right)^2/4 
+n^{\pm }_{\rm c}.
\end{equation}
Also the conformal weight of spin sector is given as
\begin{equation}
\Delta_{\rm s}^{\pm}(\lambda_{\rm s})
=\left(N_{\rm s}\pm D_{\rm s}\right)^2/4 
+n^{\pm }_{\rm s}.
\end{equation}
In the expression (\ref{rhorhocft}), $v_{\rm c}$ and $v_{\rm s}$ 
are the velocities of the density waves in electronic and spin sectors,  
respectively. 

In the SU(2) CS model, 
these quantities turn into $v_{\rm c}=v_{\rm s}=v_{\rm F}$.\cite{HaHaldane} 
The Fermi wave number $k_{\rm F}$ is given by $\pi \rho_0/2$. 
We find that the expression (\ref{rhorhoasym}) 
has the form of eq. (\ref{rhorhocft}), considering that 
\begin{eqnarray*}
\lefteqn{g_{\rm h}^{\uparrow \uparrow}(m^2+n^2)
+ 2g_{\rm h}^{\uparrow \downarrow} m n} \\
& = &
\frac{1}{2}
\left( e^{-\alpha} 0 \pm (m + n) e^{\alpha} \right)^2 +
\frac{1}{2}
\left( 0\pm (m - n) \right)^2 \\
& = & 
2 \Delta_{\rm c}^\pm 
\left(N_{\rm c}=0, D_{\rm c}=(m+n),n_{\rm c}^\pm=0\right) 
+ 2 \Delta_{\rm s}^{\pm}
\left(N_{\rm s}=0,D_{\rm s}=(m-n),n_{\rm s}^\pm=0\right).
\end{eqnarray*}
The first term in eq. (\ref{rhorhoasym}) comes from the secondary field 
($^\exists n^{\pm}_{\rm c(s)}\ne 0$) of the vacuum state 
($N_{\rm c(s)}=D_{\rm c(s)}=0$). 
The second term originates from the primary field 
($^\forall n_{\rm c(s)}^\pm=0$). 
We thus confirm the consistency of our result on the 
dynamical density-density correlation functions 
with CFT. 
\section{SU($K$) Calogero-Sutherland model }

So far we have considered the SU(2) CS model. 
It is straightforward to generalize our results to the SU($K$) model; 
the thermodynamics has been formulated\cite{fesmsm}
and hence we can obtain the knowledge about the elementary excitations. 
The Hamiltonian of the SU($K$) model with $K\geq3$ 
is the same as (\ref{hamilmulti}). 
In this case, however, we consider particles 
with SU($K$) color $\sigma\in\{1, 2, \cdots, K\}$. 
Therefore the operator $P_{ij}$ should be regarded as 
the SU($K$) color exchange operator. 
 
Following ref. \cite{fesmsm}, thermodynamics of the SU($K$) model 
is given by
\begin{equation}
\Omega
=
-T\int_{-\infty}^{\infty}\frac{{\rm d}v}{2\pi}
\sum_{\sigma=1}^K\ln\left(1+\omega_{\sigma} ^{-1}\right),
\label{Omegak}
\end{equation}
where $\omega_{\sigma}$ is the real solution of the following equations:
\begin{equation}
\epsilon_{{\rm p}\sigma}/T
\equiv\left(v^2/2-\zeta_\sigma\right)/T
=\ln\left(1+\omega_\sigma\right)
-\sum_{\sigma'=1}^K g_{\rm p}^{\sigma,\sigma'}
\ln\left(1+\omega_{\sigma'}^{-1}\right), \ \ 
(\sigma=1, 2, \cdots, K).
\end{equation}
Here $\zeta_\sigma$ represents the chemical potential 
of each species. The expression of the $K\times K$ valued matrix  
${\bf g}_{\rm p}=(g_{\rm p}^{\sigma \sigma'})$, 
which represents the exclusion statistics of particles, is given by
\begin{equation}
g_{\rm p}^{\sigma \sigma'}=\delta_{\sigma,\sigma'}+\beta.
\end{equation}
The expression is also described as
\begin{equation}
\epsilon_{{\rm h}\sigma}/T
\equiv
\left(-v^2/(2(K\beta+1))-\zeta_{{\rm h}\sigma}\right)/T
=\ln\left(1+\omega_\sigma^{-1}\right)
-\sum_{\sigma'=1}^K g_{\rm h}^{\sigma,\sigma'}
\ln\left(1+\omega_{\sigma'}\right), \ \  
(\sigma=1, 2, \cdots, K),
\end{equation}
with 
$\zeta_{{\rm h}\sigma}
=-\zeta_{\sigma}+(\beta/(K\beta+1))\sum_{\sigma'=1}^{K}\zeta_{\sigma'}$
and 
$g_{\rm h}^{\sigma \sigma'}=\delta_{\sigma,\sigma'}-\beta/(K\beta+1)$. 

At $T=0$ and $\zeta_{\rm \sigma}=\zeta$ 
({\it i.e.}, each species takes the same chemical potential), 
the velocity distribution functions are given by
\begin{equation}
\left\{
\begin{array}{lll}
\rho_\sigma(v)=1/(K\beta +1), &
\rho^*_{\sigma}(v)=0,   \quad &
\mbox{for}\ \ 
\left|v\right|<v_{\rm F}\equiv\left(2\zeta\right)^{1/2},\\
\rho_{\sigma}(v)=0,           &
\rho^*_{\sigma}(v)=1,   \quad &
\mbox{for}\ \ 
\left|v\right|>v_{\rm F}.
\end{array}
\right.
\end{equation}
The chemical potential $\zeta$ is obtained with the relation
\begin{equation}
\int_{-\infty}^{\infty}\frac{{\rm d}v}{2\pi}
\sum_{\sigma=1}^K \rho_{\sigma}\left(v\right)
=\frac{N}{L}\equiv\rho_0
\end{equation}
as $\zeta=v_{\rm F}^2/2=\left[\pi \rho_0 \left(K\beta+1\right)/K\right]^2/2$. 

For $\left|v\right|>v_{\rm F}$, 
excitations are particle-like; quasiparticles with energy 
$\epsilon_{{\rm p}\sigma }$, charge $+1$, color $\sigma$, 
and statistics ${\bf g}_{\rm p}$. 
For $\left|v\right|<v_{\rm F}$, 
on the other hand, excitations are hole-like; quasiholes with energy 
$\epsilon_{{\rm h}\sigma}$, charge $-1/(K\beta+1)$, color $\sigma$, 
and statistics ${\bf g}_{{\rm h}}$. 
We list up datum of elementary excitations in table \ref{fulltable:2}. 

Now we consider the dynamical density-density correlation function: 
$\langle \rho(x,t)\rho\rangle$. 
First we obtain the ^^ ^^ minimal bubble", 
which is the set of charge and SU($K$) color neutral excited states 
containing the minimal number of excitations. 
The minimal bubble in this model is given by the simultaneous excitations of
\begin{equation}
\label{minimal}
\left\{
\begin{array}{l} 
\mbox{one quasiparticle with color }   \sigma  \\
\beta+1 \mbox{ quasiholes with color } \sigma \\
\mbox{$K-1$ sets of $\beta$ quasiholes with colors }   \sigma'
\in \{1, 2, \cdots, K\}\setminus \{\sigma\}.
\end{array}
\right.
\end{equation}
The neutrality with respect to both charge and SU($K$) colors is verified by 
$\zeta_{\sigma}
+\left(\beta +1\right)\zeta_{{\rm h}\sigma}
+\sum_{\sigma'(\ne \sigma)}\beta\zeta_{{\rm h}\sigma'}=0$. 
Notice that the assignment of colors for the quasiholes
is not the same as SU(2) case.
Following the method in the previous section 
with table \ref{fulltable:2} and (\ref{minimal}), 
we arrive at the following expression for 
$\langle\rho(x,t)\rho\rangle$:
\begin{eqnarray}
\langle \rho(x,t)  \rho \rangle
& = & 
C
\int_{|u| \geq v_{\rm F}} {\rm d} u
\prod_{i=1}^{\beta+1}
\int_{|v_i| \leq v_{\rm F}} {\rm d} v_i
\prod_{\alpha}^{K-1}
\prod_{j=1}^{\beta}
\int_{|w_j| \leq v_{\rm F}} {\rm d} w_j^{\alpha}
\exp [{\rm i}(Q x - E t)]Q^2 \nonumber\\
& & 
\times\frac{
   \prod_{i < j}^{\beta+1}
       (v_i - v_j )^{2 g_{\rm d}}
   \prod_{\alpha}^{K-1}
   \prod_{i < j}^{\beta}
       (w_i^{\alpha} - w_j^{\alpha} )^{2 g_{\rm d}} }
{
      [ \; \epsilon_{\rm p}(u) \; ]^{1-g_{\rm p}}
      [ \; \prod_{i=1}^{\beta+1}
            \epsilon_{\rm h}(v_i)
           \prod_{\alpha}^{K-1}
           \prod_{j=1}^{\beta}
            \epsilon_{\rm h}(w_j^{\alpha}) 
        \; ]^{1-g_{\rm d}}}
       \nonumber\\
& &
\times\frac{
   \prod_{\alpha}^{K-1}
   \prod_{i=1}^{\beta+1}
   \prod_{j=1}^{\beta}
   (v_i - w_j^{\alpha})^{2 g_{\rm o}}
   \prod_{\alpha < \alpha' }^{K-1}
   \prod_{i,j=1}^{\beta}
       (w_i^{\alpha} - w_j^{\alpha'})^{2 g_{\rm o}} }
     { \prod_{i=1}^{\beta+1}
       (u - v_i)^2 }, 
\label{rhorhosuk}
\end{eqnarray}
where
$$
g_{\rm p} = g_{\rm p}^{\sigma,\sigma}=\beta+1 , \;
g_{\rm d} = g_{\rm h}^{\sigma,\sigma}=\frac{(K-1)\beta+1}{K \beta+1} , \;
g_{\rm o} = g_{\rm h}^{\sigma,\sigma'(\ne \sigma)}=\frac{-\beta}{K
\beta+1}.$$ The subscripts of ^^ ^^ d" and ^^ ^^ o" represent 
the diagonal and off-diagonal part, respectively. 
Energies and momenta are given by
\begin{eqnarray}
E&=&\epsilon_{{\rm p}}(u)
+\sum_{i=1}^{\beta +1}\epsilon_{{\rm h}}(v_i)
+\sum_{\alpha}^{K-1}\sum_{j=1}^\beta \epsilon_{{\rm h}}(w_j^\alpha),
\\
Q&=&u
-\frac{1}{K\beta +1}\left(
\sum_{i=1}^{\beta+1}v_i
+\sum_{\alpha}^{K-1}\sum_{j=1}^{\beta}w_{j}^\alpha
\right). 
\end{eqnarray}

The factor $C$ is obtained by the charge susceptibility
\begin{equation}
\chi=\frac{K^2}{\pi^2 \rho_0 \left(K\beta +1\right)^2}
\end{equation}
and the Kramers-Kronig relation in the same way as the SU(2) case; 
the expression for $C$ is given as
\begin{equation}
C=\frac{K}{8\pi^2 \left(K\beta +1\right)^{\beta +1}S(\beta;K)}.
\end{equation}
Here $S(\beta;K)$ is given by an integral
analogous to the Selberg integral:
\begin{eqnarray}
 S(\beta;K) & = &
      \prod_{i=1}^{\beta+1} 
           \int_0^1 {\rm d} p_i
      \prod_{\alpha}^{K-1} 
      \prod_{j=1}^{\beta} 
           \int_0^1 {\rm d} q_j^{\alpha}
 \prod_{i=1}^{\beta+1}
   \left[
      p_i (1 - p_i)
   \right]^{g_{\rm o}}
 \prod_{\alpha}^{K-1}
 \prod_{j=1}^{\beta}
    \left[
      q_j^{\alpha} (1 - q_j^{\alpha})
   \right]^{g_{\rm o}} 
\nonumber\\
           &  &
     \times
  \prod_{i<j}^{\beta+1} 
        (p_i-p_j)^{2g_{\rm d}}
  \prod_{\alpha}^{K-1} \prod_{i<j}^{\beta} 
        (q_i^{\alpha}-q_j^{\alpha})^{2g_{\rm d}}
  \prod_{\alpha}^{K-1}
  \prod_{i=1}^{\beta+1}
  \prod_{j=1}^{\beta}
       (p_i - q_j^{\alpha})^{2g_{\rm o}}
  \prod_{\alpha < \alpha'}^{K-1}
  \prod_{i,j=1}^{\beta}
       (q_i^{\alpha} - q_j^{\alpha'})^{2g_{\rm o}}.
\nonumber\\
\end{eqnarray}

Similarly with the SU(2) model, we can confirm the consistency 
of the expression (\ref{rhorhosuk}) with CFT.  
In the asymptotic region, the expression (\ref{rhorhosuk}) turns into 
\begin{equation}
\langle \rho (x,t)\rho\rangle
\sim 
A\left(\frac{1}{\xi_{\rm R}^2} 
+\frac{1}{\xi_{\rm L}^2}\right)
+\sum \frac{A_{m,\left\{m_\alpha\right\}}
\cos\left[2k_{\rm F}x\left(m+\sum_{\alpha}^{K-1}m_\alpha\right)\right]}
{\left(\xi_{\rm R}\xi_{\rm L}\right)^{2\Delta\left[m,\left\{m_\alpha\right\}
\right]}},
\end{equation}
with $k_{\rm F}=\pi \rho_0/K$ and 
\begin{equation}
\Delta\left[m,\left\{m_\alpha\right\}\right]
=\frac12\left(m^2 +\sum_{\alpha}^{K-1}m_\alpha^2\right)
+\frac12 \left(m+\sum_{\alpha}^{K-1}m_\alpha\right)^2 g_{\rm o}.
\label{rhorhoasymsuk}
\end{equation}
Here $A$ and $A_{m,\left\{m_\alpha\right\}}$ are constants,
and we put $\xi_{\rm R}=x-v_{\rm F}t$ and $\xi_{\rm L}=x+v_{\rm F}t$. 
The integers $m$ and $m_{\alpha}$ represent the number of quasiholes 
with each color near a specified Fermi point: 
when the quasiparticle is near $u=v_{\rm F}$, $m$ 
quasiholes are near $v_i=-v_{\rm F}$ with color $\sigma$ 
and $m_\alpha$ quasiholes with color $\alpha$ 
are near $w_j^\alpha =-v_{\rm F}$; 
when the quasiparticle is near $u=-v_{\rm F}$, 
on the other hand, $m$ quasiholes with color $\sigma$ distribute 
near $v_i=v_{\rm F}$ and $m_\alpha$ quasiholes with color $\alpha$ distribute 
near $w_j^\alpha=v_{\rm F}$. 
The integers $m$ and $m_\alpha$ run over 
$\left\{0, 1,2,\cdots,\beta+1\right\}$ 
and $\left\{0, 1,2,\cdots,\beta\right\}$, respectively.

For the SU($K$) CS model, 
we can obtain the conformal weight of CFT from ref. \cite{Kawakami} as
\begin{equation}
\Delta^\pm\left[{\bf n,\bf j}\right]
=\frac18 {\bf n}^t {\bf g}_{\rm p}{\bf n}
+\frac12 {\bf j}^t {\bf g}_{\rm h}{\bf j}+n^{\pm}.
\end{equation}
Here $\Delta^+$ $(\Delta^-)$ represents the conformal weight of 
holomorphic (antiholomorphic ) part. 
The $K$ component vectors ${\bf n}$ 
and ${\bf j}$ specify the primary field; 
the $\sigma$-th component $n_\sigma$ in ${\bf n}$ 
represents the change of number of particles with color $\sigma$. 
The component $j_\sigma$ in ${\bf j}$ represents 
the current $2k_{\rm F}j_\sigma$ excitation carried by particles 
with color $\sigma$.
The non-negative integers $n^\pm$ represent the secondary field contributions.

In the density-density correlation function, 
only the excited states with ${\bf n}=0$ is relevant. 
If we identify ${\bf j}$ as $\left(\left\{m_\alpha\right\},m\right)$, 
we can see that the expression (\ref{rhorhoasymsuk}) 
is consistent with CFT; 
The first term in the right-hand side of (\ref{rhorhoasymsuk}) 
comes from the secondary fields $\left(n^+,n^- \right)=(1,0)$ 
or  $(0,1)$ of the vacuum state ${\bf j}=0$. 
The rest of the contribution comes from the primary fields 
with ${\bf j}\ne 0$.

\section{Conclusion and Discussion}

We have proposed an elementary method of constructing  
the dynamical correlation functions of the SU($K$) CS model. 
On the other hand,
hole part of the one-particle Green function of the SU(2) CS model 
has been obtained in ref. \cite{Green} from the finite-size calculation. 
The resultant expression can be derived by the approach in the present paper.
All results can be summarized in simple formulae 
for the density-density correlation 
function $\langle \rho(x,t)  \rho \rangle$ 
and hole propagator $G(x,t)$;
\begin{eqnarray}
\langle \rho(x,t)  \rho \rangle
& = & {\cal I}(1)[\ Q\ ],\\
G(x,t)
& = & {\cal I}(0)[\ 1\ ],
\end{eqnarray}
where 
\begin{eqnarray}
{\cal I}(m)[\ *\ ]=
&&
\prod_{i=1}^{m}
\int_{|u_i| \geq v_{\rm F}} {\rm d} u_i
\prod_{j=1}^{\beta+1}
\int_{|v_j| \leq v_{\rm F}} {\rm d} v_j
\prod_{\alpha}^{K-1}
\prod_{k=1}^{\beta}
\int_{|w_k| \leq v_{\rm F}} {\rm d} w_k^{\alpha}
\exp [{\rm i}(Q x - E t)]                    
\nonumber\\
&& \ \ \ \ \ \ \ \ \ \ \ \ \ \ \ \ \ \ \ \ \ \ \ \ \ \ \ 
\quad\qquad
\times 
[\ *\ ]^2
F(m|\left\{u_i\right\},\left\{v_j\right\},\left\{w_k\right\}),
\end{eqnarray}
with 
\begin{eqnarray}
F(m|\left\{u_i\right\},\left\{v_j\right\},\left\{w_k\right\})=
&&C(m)
\frac{
   \prod_{i < j}^{\beta+1}
       (v_i - v_j )^{2 g_{\rm d}}
   \prod_{\alpha}^{K-1}
   \prod_{i < j}^{\beta}
       (w_i^{\alpha} - w_j^{\alpha} )^{2 g_{\rm d}} }
{
      [ \; \prod_{i=1}^{m}\epsilon_{\rm p}(u_i) \; ]^{1-g_{\rm p}}
      [ \; \prod_{i=1}^{\beta+1}
            \epsilon_{\rm h}(v_i)
           \prod_{\alpha}^{K-1}
           \prod_{j=1}^{\beta}
            \epsilon_{\rm h}(w_j^{\alpha}) 
        \; ]^{1-g_{\rm d}}}
       \nonumber\\
& &
\times\frac{
   \prod_{\alpha}^{K-1}
   \prod_{i=1}^{\beta+1}
   \prod_{j=1}^{\beta}
   (v_i - w_j^{\alpha})^{2 g_{\rm o}}
   \prod_{\alpha < \alpha' }^{K-1}
   \prod_{i,j=1}^{\beta}
       (w_i^{\alpha} - w_j^{\alpha'})^{2 g_{\rm o}} }
     { \prod_{i=1}^{m}\prod_{j=1}^{\beta+1}
       (u_i - v_j)^2 }. 
\end{eqnarray}
Here $C(m)$ is some constant.
The important point to note is that (non-trivial part of) 
the form factor can essentially be characterized 
by the statistical interactions.

Although we studied only the integer $\beta$ case,
it is straightforward to generalize the rational $\beta$ case,
in the same way as the spinless case.\cite{Hal-rev,Ha}
We speculate that our method 
is applicable to the density-density correlation function 
and hole part of the Green function 
of the CS model with arbitrary internal symmetry. 

Finally, we discuss the applicability of our approach 
to the lattice versions of the CS models: 
Haldane-Shastry (HS) model,\cite{HS1,HS2} 
supersymmetric $1/r^2$ $t${\it -}$J$ model\cite{KY} 
and their multicomponent models\cite{HaHaldane,Kawakami}. 
The HS model and $1/r^2$ $t$-$J$ model 
are thermodynamically equivalent to systems of free particles 
obeying fractional exclusion statistics.\cite{KK} 
Thus we expect that our unified description of thermodynamics 
and dynamics is applicable to the two models. 
Actually we note that the known results \cite{HZ} 
on the dynamics of HS model can be interpreted in the same way as \S 2-4. 

The thermodynamics of SU($K$) HS models ($K>2$),\cite{KK} 
on the other hand, cannot be described 
by the fractional exclusion statistics in the unpolarized case; 
SU($K$) HS model ($K>2$) is thermodynamically equivalent to 
the system of free {\it parafermion} of order $K-1$. 
Hence the dynamics of the SU($K$) HS model ($K>2$) 
is beyond the applicability of our approach. 
The relation between dynamics and thermodynamics 
of free parafermion is another interesting problem.

\section*{Acknowledgments}
The authors thank Y. Kuramoto for valuable discussions. 
Y. K. was supported by a Grant-in-Aid for Encouragement of 
Young Scientists (08740306) from 
the Ministry of Education, Science, Sports and Culture of Japan. 


\begin{table}
\caption{Physical properties of elementary excitations
of the SU(2) CS model.}
\label{fulltable:1}
\begin{tabular}{@{\hspace{\tabcolsep}\extracolsep{\fill}}cccccc} 
\hline
species & charge  & spin & energy & momentum & statistics \\ 
\hline
quasiparticle $\sigma$  & 1&  $\sigma$ &
$v^2/2-\zeta-\sigma h$& $v$& ${\bf g}_{\rm p}$ \\
quasihole $\sigma$  & $-1/(2\beta+1)$  &$-\sigma$ 
&$\left(\zeta-v^2/2\right)/(2\beta+1)+\sigma h$ 
&$-v/(2\beta+1)$ &${\bf g}_{\rm h}$ \\
\hline
\end{tabular}
\end{table}

\begin{table}
\caption{Physical properties of elementary excitations of 
the SU($K$) CS model.}
\label{fulltable:2}
\begin{tabular}{@{\hspace{\tabcolsep}\extracolsep{\fill}}cccc} 
\hline
species & charge  & energy & statistics \\ 
\hline
quasiparticle $\sigma$ &1 
&$v^2/2-\zeta_\sigma$                       & ${\bf g}_{\rm p}$\\
quasihole $\sigma$     &$-1/(K\beta+1)$
&$-v^2/(2(K\beta+1))-\zeta_{{\rm h}\sigma}$ & ${\bf g}_{\rm h}$\\
\hline
\end{tabular}
\end{table}


\begin{thebibliography}{99}


\bibitem{Cal}
F.~Calogero:
J. Math. Phys. {\bf 10} (1969) 2191, 2197.

\bibitem{Suth}
B.~Sutherland: 
Phys. Rev. A{\bf 4} (1971) 2019, 
A{\bf 5} (1972) 1372, 
J. Math. Phys. {\bf 12} (1971) 246, 251.


\bibitem{Haldanefrac}
F.~D.~M.~Haldane: 
Phys. Rev. Lett. {\bf 66} (1991) 1529.
 


\bibitem{SLA}
B.~D.~Simons, P.~A.~Lee and B.~L.~Altshuler: 
Phys. Rev. Lett. {\bf 70} (1993) 4122, 
{\bf 72} (1994) 64,
Phys. Rev. B {\bf 48} (1993) 11450,
Nucl. Phys. B{\bf 409} (1993) 487.


\bibitem{HZ}
F.~D.~M.~Haldane and M.~R.~Zirnbauer:
Phys. Rev. Lett. {\bf 71} (1993) 4055.


\bibitem{Ha}
Z.~N.~C.~Ha:
Phys. Rev. Lett. {\bf 73} (1994) 1574, 
{\bf 74} (1995) 620 (errata),
Nucl. Phys. B{\bf 435} (1995) 604.

\bibitem{LPS}
F.~Lesage, V.~Pasquier and D.~Serban:
Nucl. Phys. B{\bf 435} (1995) 585.


\bibitem{Forr}
P.~J.~Forrester:
Mod. Phys. Lett. B {\bf 9} (1995) 359,
J. Math. Phys. {\bf 36} (1995) 86.
 

\bibitem{Konn}
H. Konno:
J. Phys. A {\bf 29} (1996) L191,
Nucl. Phys. B{\bf 473} (1996) 579.


\bibitem{MP1}
J.~Minahan and A.~P.~Polychronakos:
Phys. Rev. B {\bf 50} (1994) 4236.

\bibitem{Khve}
D.~V.~Khveshchenko:
Int. J. Mod. Phys. B {\bf 9} (1995) 1639.

\bibitem{Tsve}
A. M. Tsvelik:
{\em Mapping of the Calogero-Sutherland Model 
to the Gaussian Model},
preprint (1996) cond-mat/9603203.



\bibitem{Hal-rev}
F.~D.~M.~Haldane:
in 
{\em Correlation Effects in Low Dimensional Electron 
Systems}, 
Springer Series in Solid-State Science {\bf 118},
A.~Okiji and N.~Kawakami (eds.)
(Springer-Verlag, 1994),
in
{\em Modern Quantum Field Theory II},
S. R. Das, G. Mandal, S. Mukhi and S. R. Wadia (eds.)
(World Scientific, 1995).



\bibitem{HaHaldane}
Z.~N.~C.~Ha and F.~D.~M.~Haldane: 
Phys. Rev. B {\bf 46} (1992) 9359.

\bibitem{MP2}
J.~A.~Minahan and A.~P.~Polychronakos:
Phys. Lett. B {\bf 302} (1993) 265.
%
\bibitem{HW}
K.~Hikami and M.~Wadati:
J. Phys. Soc. Jpn. {\bf 62} (1993) 469,
Phys. Lett. A {\bf 469} (1993) 263.


\bibitem{HS1} 
F. D. M. Haldane: 
Phys. Rev. Lett. {\bf 60} (1988) 635.

\bibitem{HS2} 
B. S. Shastry: 
Phys. Rev. Lett. {\bf 60} (1988) 639.


\bibitem{KY} 
Y. Kuramoto and H. Yokoyama:  
Phys. Rev. Lett. {\bf 67} (1991) 1338.


\bibitem{Polychronakos}
A. Polychronakos:
Phys. Rev. Lett. {\bf 70} (1993) 2329.

\bibitem{SS}
B.~Sutherland and B.~S.~Shastry:
Phys. Rev. Lett. {\bf 71} (1993) 5.

\bibitem{KaKu1}
Y.~Kato and Y.~Kuramoto: 
Phys. Rev. Lett. {\bf 74} (1995) 1222.


\bibitem{Green} 
Y. Kato:
{\em Green Function of the Sutherland Model
with SU(2) Internal Symmetry},
Phys. Rev. Lett. (to be published) cond-mat/9702003.



\bibitem{Uglov} 
D. Uglov:
{\em Yangian Gelfand-Zetlin Bases, gl(N)-Jack Polynomials 
and Computation of Dynamical Correlation Functions 
in the Spin Calogero-Sutherland Model},
preprint (1997) hep-th/9702020.



\bibitem{YY}
C. N. Yang and C. P. Yang:
J. Math. Phys. {\bf 10} (1969) 1115.


\bibitem{Wu} 
Y.-S. Wu:
Phys. Rev. Lett. {\bf 73} (1994) 922,
{\bf 74} (1995) 3906 (errata).

\bibitem{BW}
D.~Bernard and Y.-S.~Wu:
in 
{\it New Developments of Integrable Systems and Long-Ranged 
Interaction Models}, M.-L. Ge and Y.-S. Wu (eds.)
(World Scientific, 1995).


\bibitem{fesmsm} 
Y. Kato and Y. Kuramoto: 
J. Phys. Soc. Jpn. {\bf 65} (1996) 77.


\bibitem{FK1}
T. Fukui and N. Kawakami:
Phys. Rev. B {\bf 51} (1995) 5239,
J. Phys. A {\bf 28} (1995) 6027.



\bibitem{Kawakami} 
N. Kawakami: 
Phys. Rev. B{\bf 46} (1992) 1005, 3191
J. Phys. Soc. Jpn. {\bf 62} (1993) 2270, 4163.


\bibitem{TLL}
F. D. M. Haldane:
J. Phys. C {\bf 14} (1981) 2585;
%
J. Voit:
Rep. Prog. Phys. {\bf 58} (1995) 977.


\bibitem{BPZ}
A. A. Belavin, A. M. Polyakov and A. B. Zamolodchikov:
Nucl. Phys. B{\bf 241} (1994) 333.

\bibitem{Cardy}
J. L. Cardy: Nucl. Phys. B{\bf 270} (1986) 186.



\bibitem{KK} 
Y. Kuramoto and Y. Kato: 
J. Phys. Soc. Jpn. {\bf 64} (1995) 4518; 
Y. Kato and Y. Kuramoto: 
{\it ibid.} {\bf 65} (1996) 1622.


\end{thebibliography}
\end{document}